\documentclass[journal,twoside,web]{ieeecolor}
\usepackage{tmi}
\usepackage{cite}
\usepackage{amsmath,amssymb,amsfonts}
\usepackage{graphicx}
\usepackage{textcomp}
\usepackage[ruled,linesnumbered]{algorithm2e}
\usepackage{amsmath,amsfonts}
\usepackage{array}
\usepackage[caption=false,font=normalsize,labelfont=sf,textfont=sf]{subfig}
\usepackage{textcomp}
\usepackage{stfloats}
\usepackage{url}
\usepackage{verbatim}
\usepackage{graphicx}
\usepackage{cite}
\usepackage{array,booktabs}
\newcolumntype{L}{@{}>{\kern\tabcolsep}l<{\kern\tabcolsep}}
\usepackage{colortbl}
\usepackage{xcolor}
\let\labelindent\relax
\usepackage[inline]{enumitem}
\colorlet{linkequation}{blue}
\usepackage{ragged2e}
\usepackage{gensymb}
\usepackage{tabularx}
\usepackage{svg}
\usepackage[colorlinks=true, citecolor=blue, linkcolor=blue]{hyperref}
\usepackage{cleveref}

\newcommand{\removelatexerror}{\let\@latex@error\@gobble}

\def\BibTeX{{\rm B\kern-.05em{\sc i\kern-.025em b}\kern-.08em
    T\kern-.1667em\lower.7ex\hbox{E}\kern-.125emX}}
\begin{document}
\title{A Novel Training Framework for Physics-informed Neural Networks: Towards Real-time Applications in Ultrafast Ultrasound Blood Flow Imaging}
\author{Haotian Guan, Jinping Dong,~\IEEEmembership{Member,~IEEE}, Wei-Ning Lee,~\IEEEmembership{Member,~IEEE}
\thanks{This work was supported in part by Hong Kong Health and Medical Research Fund (08192616). }
\thanks{Haotian Guan is with the Department of Electrical and Electronic Engineering, The University of Hong Kong.}
\thanks{Jinping Dong was with the Department of Electrical and Electronic Engineering, The University of Hong Kong, and is now with the Department of Biomedical Engineering, and also with Beijing International Science and Technology Cooperation Base for Intelligent Physiological Measurement and Clinical Transformation, College of Chemistry and Life Science,Beijing University of Technology, Beijing 100020, China.}  
\thanks{Wei-Ning Lee is with the Department of Electrical and Electronic Engineering, The University of Hong Kong, Hong Kong 999077, China, and also with the Biomedical Engineering Programme, The University of Hong Kong, Hong Kong (e-mail: wnlee@eee.hku.hk).}}
\maketitle

\begin{abstract}
Ultrafast ultrasound blood flow imaging is a state-of-the-art technique for depiction of complex blood flow dynamics \emph{in vivo} through thousands of full-view image data (or, timestamps) acquired per second. Physics-informed Neural Network (PINN) is one of the most preeminent solvers of the Navier-Stokes equations, widely used as the governing equation of blood flow. However, that current approaches rely on full Navier-Stokes equations is impractical for ultrafast ultrasound. We hereby propose a novel PINN training framework for solving the Navier-Stokes equations. It involves discretizing Navier-Stokes equations into steady state and sequentially solving them with test-time adaptation. The novel training framework is coined as SeqPINN. Upon its success, we propose a parallel training scheme for all timestamps based on averaged constant stochastic gradient descent as initialization. Uncertainty estimation through Stochastic Weight Averaging Gaussian is then used as an indicator of generalizability of the initialization. This algorithm, named SP-PINN, further expedites training of PINN while achieving comparable accuracy with SeqPINN. The performance of SeqPINN and SP-PINN was evaluated through finite-element simulations and \emph{in vitro} phantoms of single-branch and trifurcate blood vessels. Results show that both algorithms were manyfold faster than the original design of PINN, while respectively achieving Root Mean Square Errors of 0.63 cm/s and 0.81 cm/s on the straight vessel and 1.35 cm/s and 1.63 cm/s on the trifurcate vessel when recovering blood flow velocities. The successful implementation of SeqPINN and SP-PINN open the gate for real-time training of PINN for Navier-Stokes equations and subsequently reliable imaging-based blood flow assessment in clinical practice.
\end{abstract}


\section{Introduction}
\label{sec:introduction}
\IEEEPARstart{D}{epicting} hemodynamics in the circulation system is important for diagnosis of cardiovascular diseases, such as myocardial infarction and ischemic stroke due to atherosclerosis. Blood flow velocity and blood pressure are key hemodynamic parameters and directly reveal vascular conditions. The two primary methods used to estimate blood flow velocity are Doppler-based \cite{overbeck1992vector} and speckle-tracking-based \cite{trahey1987angle}. Doppler-based methods accurately estimate the velocity component along the beam direction only, and they suffer from Doppler angle ambiguity and aliasing in fast flow phases of the cardiac cycle \cite{jensen2016ultrasound}. Speckle-tracking-based methods can estimate the two-dimensional (2D) velocity field directly. They assume that local echogenicity is well-preserved, or ultrasound signals are highly correlated between two consecutive frames, but this assumption does not hold for a high velocity blood flow. Fast flow issues have been prevalently addressed by ultrafast ultrasound with unfocused wave transmissions. Ultrafast ultrasound permits full-view data acquisitions at thousands of frames per second \cite{bercoff2011ultrafast}, but it is known to degrade image quality and introduce artifacts \cite{udesen2008high}. The multi-factorial characteristic of ultrasound blood flow velocity estimation leads to velocity estimation bias and variance.

Physics-constrained optimization has been proposed in the literature for the regularization of both Doppler-based and speckle-tracking-based blood flow velocity estimates. Muth et al. proposed a protocol called 'DeAN', which segmented the aliasing area first and then performed dealiasing and denoising on color Doppler echocardiography \cite{muth2011unsupervised}. Similarly, segmentation followed by physics-constrained optimization using planar mass conservation and free-slip boundary conditions \cite{vixege2021physics} and deep learning-based segmentation in conjunction with phase unwrapping were also proposed to solve the aliased area \cite{ling2023phase, nahas2020deep} in color Doppler. However, one limitation of this protocol is the dependency on the segmentation algorithm. Given the uncertainty in generalization of current deep learning models, a reliable performance for all situations may not be guaranteed. When it comes to ultrafast ultrasound of blood flow, physics-based regularization is challenged by a huge amount of sub-optimal quality data acquired at ultra-high frame rates. 

Since its emergence, physics-informed neural network (PINN) \cite{raissi2019physics} has made a transformational impact on solving partial differential equations (PDEs) \cite{hua2023physics, cai2021physics, oszkinat2022uncertainty}. As a new class of artificial neural networks, recent research on PINN has demonstrated its excellence in coping with imperfect situations, such as no specified initial or boundary conditions and noisy data \cite{wang2021deep, kharazmi2021hp, CiCP-28-2002}. PINN is mesh-free, making it more flexible for arbitrary geometries and more computationally efficient. PINN has also been applied extensively in medical domains. Buoso et al.\cite{buoso2021personalising} devised a left-ventricular shape model and a parametric PINN to simulate the mechanical behavior of the left ventricle, such as ejection fraction, stroke work, etc. Herten et al. \cite{van2022physics} introduced PINN to solve differential equations in tracer-kinetics for quantification of myocardial perfusion. Ferdian et al. incorporated PINN into a deep residual network to quantify intracranial blood velocity, flow, and relative pressure from cerebrovascular super-resolution 4D Flow MRI \cite{ferdian2023cerebrovascular}. Sarabian et al. designed area surrogate PINN (ASPINN) to estimate hemodynamic variables \cite{sarabian2022physics}.

Despite the success of PINN in many situations as an alternative of computational fluid dynamics (CFD) models, the training speed of PINN has not been demonstrated to be faster than CFD \cite{cai2021physics}. To achieve efficient learning of PINN, one instinct is to adopt pre-trained models. However, the use of pre-trained PINN models for ultrafast ultrasound blood flow imaging, like the Gordian knot, is tied by two aspects. First, training of PINN is associated with the domain geometry and corresponding initial and boundary conditions, so PINN models are usually hard to generalize to different geometries or fluid flow patterns \cite{kashefi2022physics}. Retraining PINN is thus imperative but unrealistic due to the large volume of ultrafast ultrasound data and time-consuming optimization process for a patient-specific model. Second, Navier-Stokes equations are highly nonlinear and require computation of second derivatives of the fluid velocity field. The loss landscape of PINN can be difficult to optimize \cite{krishnapriyan2021characterizing}. 

Abundant efforts have been made to expedite the training of PINN and can be summarized into two categories that circumvent the use of pre-trained models and another category that adopts pre-trained model with meta-learning. The first category aims to accelerate the convergence of PINN. Xiang \cite{xiang2021self} and Yu \cite{yu2022gradient} focused on alleviating optimization challenges by modifying the loss function. One study \cite{chiu2022can} proposed to use numerical differentiation coupled with automatic differentiation to enhance the reliability of derivative computation and thus resulted in fast and accurate convergence of PINN. The second category takes advantages of parallel training by decomposing the computational domain. cPINN \cite{jagtap2020conservative} offered space parallelization, while XPINN \cite{CiCP-28-2002} provided both space and time parallelization; nonetheless, dividing domains could be tricky when a complex geometry or a pulsatile velocity is associated with a blood flow profile. The third category relieves the optimization challenges of PINN by meta-learning to find better initialization for test data (or, tasks). Liu et al. \cite{liu2022novel} applied the classic Raptile framework from meta-learning to initialize the model parameters of PINN. Seo et al. \cite{seo2020physics} decomposed physics laws into a spatial derivative and a time derivative module. Then, they proposed a meta-learning approach to solve for the so-called reusable spatial derivative modules. However, the limitation of the meta-learning-based approaches is that the model trained during a meta-training phase must have learned the fundamental rule of finding a solution \cite{finn2017model}. Moreover, most meta-learning methods mentioned previously are impractical to incorporate a large dataset as in ultrafast ultrasound.

\begin{figure*}[!t]
\centering
\captionsetup{justification=centering}
{\includegraphics[width=0.9\linewidth]{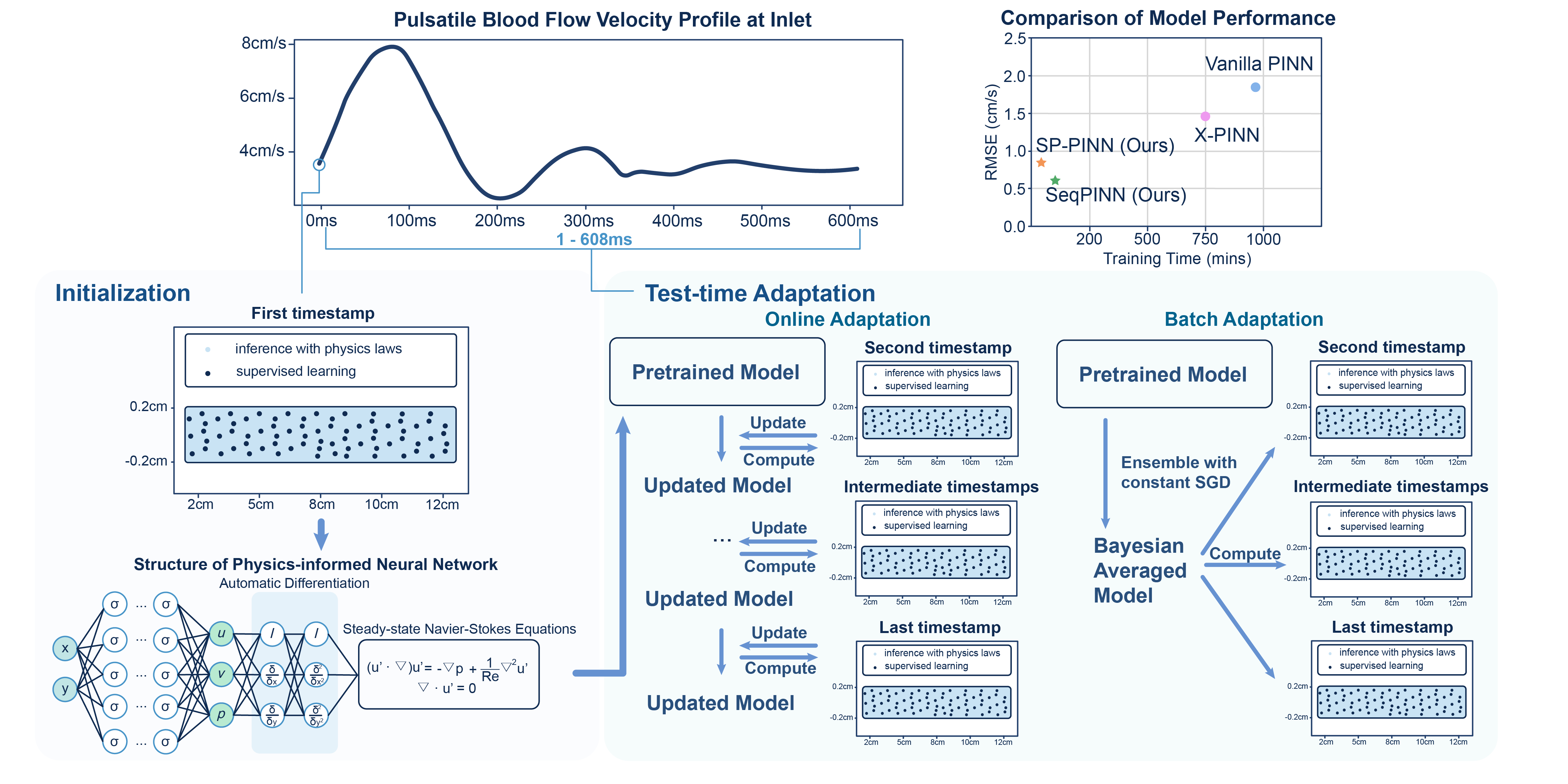}%
\label{PINN architecture}}
\caption{Our proposed sequential and parallel training frameworks of PINN with state-state Navier-Stokes equations as an example. }
\label{Fig:1}
\end{figure*}

In this work, we propose two novel training frameworks that enhance the efficiency of PINN for Navier-Stokes equations and illustrate the effectiveness of the methods with an application on ultrafast ultrasound blood flow imaging. Specifically, we first demonstrate that PINN is capable of solving Navier-Stokes equations under the steady-state assumption. Then, we initialize the model with the solution of steady-state Navier-Stokes equations and adopt an Online Test-time Adaptation (OTTA) strategy for subsequent frames. Test-time Adaptation (TTA) aims to adapt a pre-trained model from a source domain to a target domain before making predictions. In general, TTA methods can be divided into three categories by the property of test data: test-time domain adaptation, test-time batch adaptation (TTBA), and online test-time adaptation (OTTA) \cite{liang2024comprehensive}. The current PINN training framework exhibits potential to achieve fast updates because of its lightweight design of the architecture, a low dimensional data input, and PDE-guided loss functions. These advantages ensure TTA to be efficient when applied to PINN. Besides, despite being  possibly noisy, the sparse data (data and label pairs) involved in solving the forward problem with PINN are typically very difficult to obtain TTA setting. The joint use of PINN and TTA significantly reduces the optimization challenge and computational burden of training and adaptation. The framework, coined as SeqPINN, is illustrated in the flowchart in Fig. \ref{Fig:1}. With a further assumption of independent timestamps, we propose Sampled-Posterior PINN, coined as SP-PINN. As shown in Fig. \ref{Fig:1}, SP-PINN solves Navier-Stokes equations with PINN in a TTBA manner.

\begin{enumerate}
\item 
This is the first work of PINN dedicated to guiding the direct correction of blood flow velocities estimated under ultrafast ultrasound by principles of physics. This is a significantly enhanced version from our arXiv version \cite{guan2023towards}.   
\item 
We propose a fundamentally new way to adopt pre-trained models for PINN by incorporating test-time adaptation.  To the best of our knowledge, this is the first work that expands the landscape of PINN to test-time adaptation.
\item 
SeqPINN and SP-PINN are of high practical value since they efficiently provide physics-regularized blood flow estimates from ultrafast ultrasound in clinical settings. The proposed algorithms show good applicability as they can be implemented either on CPUs, which are widespread, or on GPUs, which enable even faster training. SeqPINN and SP-PINN also have the merit of being incorporated with many existing acceleration techniques on PINN.
\end{enumerate}
\section{Methods}
\subsection{Blood Flow in Large Arteries} \label{assumption}
Flow dynamics is mostly governed by Navier-Stokes equations as follows: 

\begin{align}
\begin{split}
\label{eq:3}
\frac{\partial \mathbf{u}}{\partial t}+(\mathbf{u} \cdot \nabla) \mathbf{u} &=-\frac{1}{\rho} \nabla p+ \nu \nabla^{2} \mathbf{u} \\
\nabla \cdot \mathbf{u} &=0, 
\end{split}
\end{align}
where $\mathbf{u}$ is the velocity vector of the fluid, $p$ is the pressure, $\rho$ is the density, $\nu$ is the kinematic viscosity, and $\nabla$ is the gradient differential operator. 
In this work, we assume that the blood flow starts at a steady state, and each timestamp follows the steady state with an infinitesimal change in flow velocity and luminal pressure. Theoretically, a steady state is achieved when \begin{enumerate*}[label=(\roman*)] 
\item all dependent variables of Navier-Stokes equations are independent of time, and \item the Reynolds number is sufficiently small \cite{john2016finite} \end{enumerate*}. Time independence can be equivalent to a constant flow and a constant pressure at a spatial location. To fulfill this requirement, we simulate infinitesimal time steps, making the change in blood velocity and pressure across time approach zero. A small Reynolds number, $Re$, can characterize a fully developed flow when the flow velocity is small. The loss function of PINN for non-dimensionalized steady-state Navier-Stokes equations can be written as 
\begin{align}
(\mathbf{u^{\prime}} \cdot \nabla) \mathbf{u^{\prime}} &=-\nabla p+\frac{1}{\operatorname{Re}} \nabla^{2} \mathbf{u^{\prime}} \label{eq:7}\\
\nabla \cdot \mathbf{u^{\prime}} &=0, \label{eq:8}
\end{align}
where $\mathbf{u^{\prime}}$ is the non-dimensionalized velocity vector by defining $\mathbf{u^{\prime}} = \frac{u}{U}$, where $U$ is the characteristic velocity. The utilization of steady-state Navier-Stokes equations can be regarded as decomposition of the computational domain of Navier-Stokes equations, whereby the spatial and temporal modules are treated separately. The spatial module, enabling the use of pre-trained models, is shareable across timestamps since the inputs to the model are collocation points and thus identical for all timestamps.

\subsection{\textbf{Physics-informed Neural Network (PINN)}}
\subsubsection{Original Design of PINN}

The idea of solving PDEs with neural networks can be traced back to 1990s \cite{dissanayake1994neural}; however, its popularity was endowed by physics-informed deep learning in 2019 \cite{raissi2019physics}. The Vanilla PINN was first built to find solutions to nonlinear PDEs in both forward and inverse problems. PINN is outstanding due to its ability to fit experimental data while complying with any underlying laws of physics expressed by PDEs. In the case of blood flow imaging, Navier-Stokes equations, comprised of conservation of mass and conservation of momentum for Newtonian Fluids respectively as in \cref{eq:7,eq:8}, are embedded as the residual loss denoted as $\mathcal{L}_{\mathcal{F}}(\theta)$ in the loss function. Initial boundary conditions and sparse velocity samples are enforced by supervised learning using MSE (mean square error) losses, denoted by $\mathcal{L}_{\mathcal{B}}(\theta)$ and $\mathcal{L}_{\text {data }}(\theta)$, respectively. Using $\theta$ to refer the parameters in neural network, the objective of PINN can be formed as finding the best $\theta^*$ that minimizes the total loss: 
\begin{align}
\label{eq:4}
\theta^{*}=\underset{\theta}{\arg \min }\left( \mathcal{L}_{\mathcal{F}}(\theta)+ \mathcal{L}_{\mathcal{B}}(\theta)+ \mathcal{L}_{\text {data }}(\theta)\right) 
\end{align}

The structural design of PINN is usually very simple: few compositions of fully-connected layers followed by element-wise nonlinear functions to produce the output $h_k$ after each layer as in \cref{eq:2}: 
\begin{align}
        \label{eq:2}
        \mathbf{h}_{k}=\sigma\left(\mathbf{W}_{k-1}^{\top} \mathbf{h}_{k-1}+\mathbf{b}_{k-1}\right),
\end{align}
where $\sigma$ is the activation function, $k$ is the layer number, and $W_{k-1}$ and $b_{k-1}$ are the weight and bias in the corresponding fully-connected layer, respectively. The transpose of the matrix is denoted by ${\top}$. Partial derivatives are calculated using Automatic Differentiation \cite{paszke2017automatic}, which uses exact expressions with floating-point values, thus avoiding approximation errors \cite{cuomo2022scientific}. A more complex design of PINN architecture is to implement attention mechanism \cite{bahdanau2014neural} on top of the original PINN. We find that attention mechanism stabilizes the training of PINN, so we adopt it in the architecture design. Since the acronym PINN could be misleading, we use \textit{PINN} to refer to the architecture design and \textit{Vanilla PINN} to specifically refer to the original design of PINN. 
        

 \begin{figure}[!t]
\centering
\includegraphics[width=0.9\linewidth]{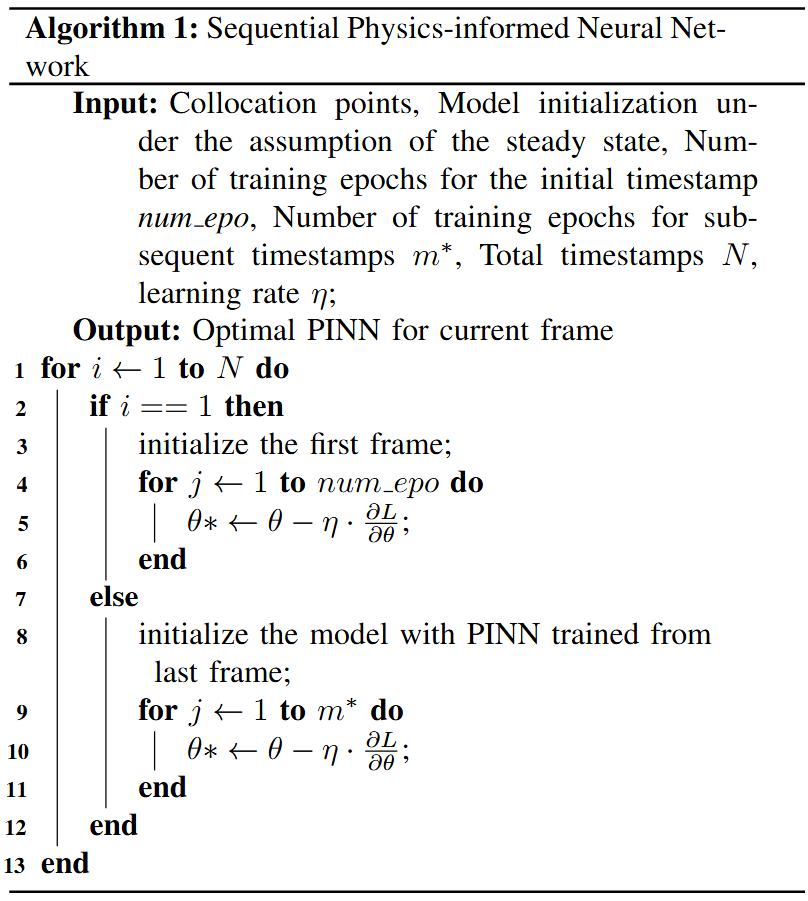}\label{alg1}
\end{figure}


\subsubsection{\textbf{SeqPINN -- Sequential Training of PINN}}

We first present an algorithm Sequential PINN, coined as SeqPINN, for learning solutions to steady-state Navier-Stokes equations and fast test-time adaptation to subsequent timestamps. SeqPINN is initialized by finding a solution for steady-state Navier-Stoke equations. The initial frame is chosen by selecting the timestamp associated with the lowest blood flow velocity in a cardiac cycle based on the temporal profile of the blood flow velocity. This prevents violation of steady-state Navier-stokes equations. Although it is possible to start from an arbitrary timestamp, we begin with end diastole, which can be determined from the R wave in an electrocardiogram (ECG). After initialization, the current PINN solution is deemed as a pre-trained model for the next timestamp. Given the boundary conditions and data points sampled at next timestamp, stochastic gradient descent (SGD) can adapt to a new local minimum with $m^{*}$ epochs, guided by the infinitesimal change in inter-timestamp velocity measurements. \Cref{III} will show that a pre-defined parameter $m^{*}$ exists that optimizes the trade-off between accuracy and training speed. In practice, $m^{*}$ can be determined by EarlyStopping. By removing the time dimension from input, we keep the spatial collocation points the same across all timestamps. This significantly reduces computational burden of SGD. The training process of SeqPINN is summarized in Algorithm 1.  

SeqPINN essentially adapts to a new model in an online manner before making predictions of the flow velocity and blood pressure at the current timestamp. Each timestamp is only observed once by the model, and previously learned knowledge can facilitate the adaptation of current model. Here, we only present the most basic version of SeqPINN. If needed, the weight of the data at a particular timestamp can be tuned by modifying the number of epochs trained at a given timestamp.
\subsubsection{\textbf{Sampled-Posterior PINN -- Parallel Training of PINN}}
Although SeqPINN significantly expedites the training of PINN while maintaining satisfactory accuracy, the OTTA scheme impedes real-time training of PINN in ultrafast ultrasound blood flow imaging. A parallel training scheme fits the goal of real-time training to the utmost. In this section, we demonstrate how we achieve parallel training across timestamps based on SeqPINN initialization and adopt the idea of SWAG \cite{maddox2019simple} to ensure trustworthy initialization.

We first define that each timestamp is an independent draw from a distribution comprised of velocity maps in a cardiac cycle. The goal is to build a pre-trained model that generalizes to all timestamps accurately with test-time adaptation using a small dataset drawn from the distribution. Similar to what we do in SeqPINN, we initialize the PINN model under the assumption of steady-state Navier-Stokes equations. Then, we view the initialization in SeqPINN $p(\theta)$ as a prior belief, and the sparse data at the next $k$ timestamps $p(\mathcal{D}\mid \theta)$ as new observations drawn from the distribution. These observations can be used to update the prior belief. As a result, a training dataset $\mathcal{D}$ comprised of $k$ timestamps is formed, and it enables Bayesian approaches to boost network accuracy and achieve better generalizability \cite{izmailov2018averaging, wilson2020bayesian, mandt2017stochastic}. The posterior belief $p(\theta\mid \mathcal{D})$ can be derived from Bayes' rule.
\begin{align}
        p(\theta\mid \mathcal{D}) = \frac{p(\mathcal{D}\mid \theta) p(\theta)}{p(\mathcal{D})},
\end{align}
where $p(\mathcal{D})$ is obtained by marginalizing $\theta$ from $p(\mathcal{D},\theta)$, which is the joint probability function of $\mathcal{D}$ and $\theta$:
\begin{align}
        p(\mathcal{D}) = \int p(\mathcal{D}, \theta ) \,d\theta, 
\end{align}
 
However, due to infinite possibilities in neural network training, the marginalization is not tractable. Alternatively, we propose a deep ensemble model with constant SGD for approximating the Bayesian posterior distribution. As proven in \cite{mandt2017stochastic}, constant SGD can be used to simulate Markov chain with a stationary distribution as an approximation for Bayesian posterior inference \cite{mandt2017stochastic}. Thus, by sequentially training the next $k$ timestamps for $m^{*}$ epochs with constant SGD, we obtain $k$ samples from the posterior distribution. Given the causality in physical systems \cite{wang2022respecting}, $m^{*}$ epochs are trained for a single timestamp (data) before the next timestamp is trained, instead of cyclically training $k$ timestamps for $m^{*}$ epochs. Finally, by averaging samples from the posterior distribution as in \cref{eq:12}, we derive a set of PINN parameters that achieves best generalizability for all timestamps.  
\begin{align}
\theta^*=\frac{1}{k} \sum_{i=1}^{k} \theta_{i} \label{eq:12}
\end{align}

Since the objective function of our neural network is non-convex, one concern is that the initialization frame may not find a flat and wide local minimum on the loss surface. This may cause the solutions of the next $k$ timestamps to have a large variance. The averaged solution may fall into the territory between two local minima and have poor generalizability. To address this, we propose an uncertainty estimation algorithm inspired by Stochastic Weight Averaging Gaussian (SWAG)\cite{maddox2019simple}, which uses the Stochastic Weight Average (SWA) solution.  SWA has been shown to improve generalizability of deep learning models without increasing inference time. It uses a single model and subtly performs an average of weights traversed by SGD iterates \cite{ruder2016overview}. SWAG forms a Gaussian distribution to approximate the posterior distribution over neural network weights. The mean $\bar{\theta}$ is the averaged SGD solutions as shown in \cref{eq:5}, and the variance of the Gaussian distribution $\Sigma$ is approximated by the solution of $k$ timestamps as shown in \cref{eq:6}, 
\begin{align}
    \bar{\theta}=\frac{1}{k} \sum_{i=1}^{k}\theta_{i} \label{eq:5}\\
    \Sigma=\frac{1}{k-1} \sum_{i=1}^{k}\left(\theta_{i}-\bar{\theta}_{i}\right)\left(\theta_{i}-\bar{\theta}_{i}\right)^{\top}, \label{eq:6}
\end{align}
where $k$ is the total number of SGD iterates and $\top$ is the matrix transpose operation. Uncertainty estimation is performed by Bayesian model averaging. The uncertainty is represented by a standard deviation map (std map) visually, and the uncertainty index is defined as the mean value in a std map. Uncertainty is monitored during training. For all timestamps, we initialize the PINN model with the averaged solution from constant SGD. This method is thus coined as Sampled-Posterior PINN (SP-PINN). 

 \begin{figure}[!t]
\centering
\includegraphics[width=0.9\linewidth]{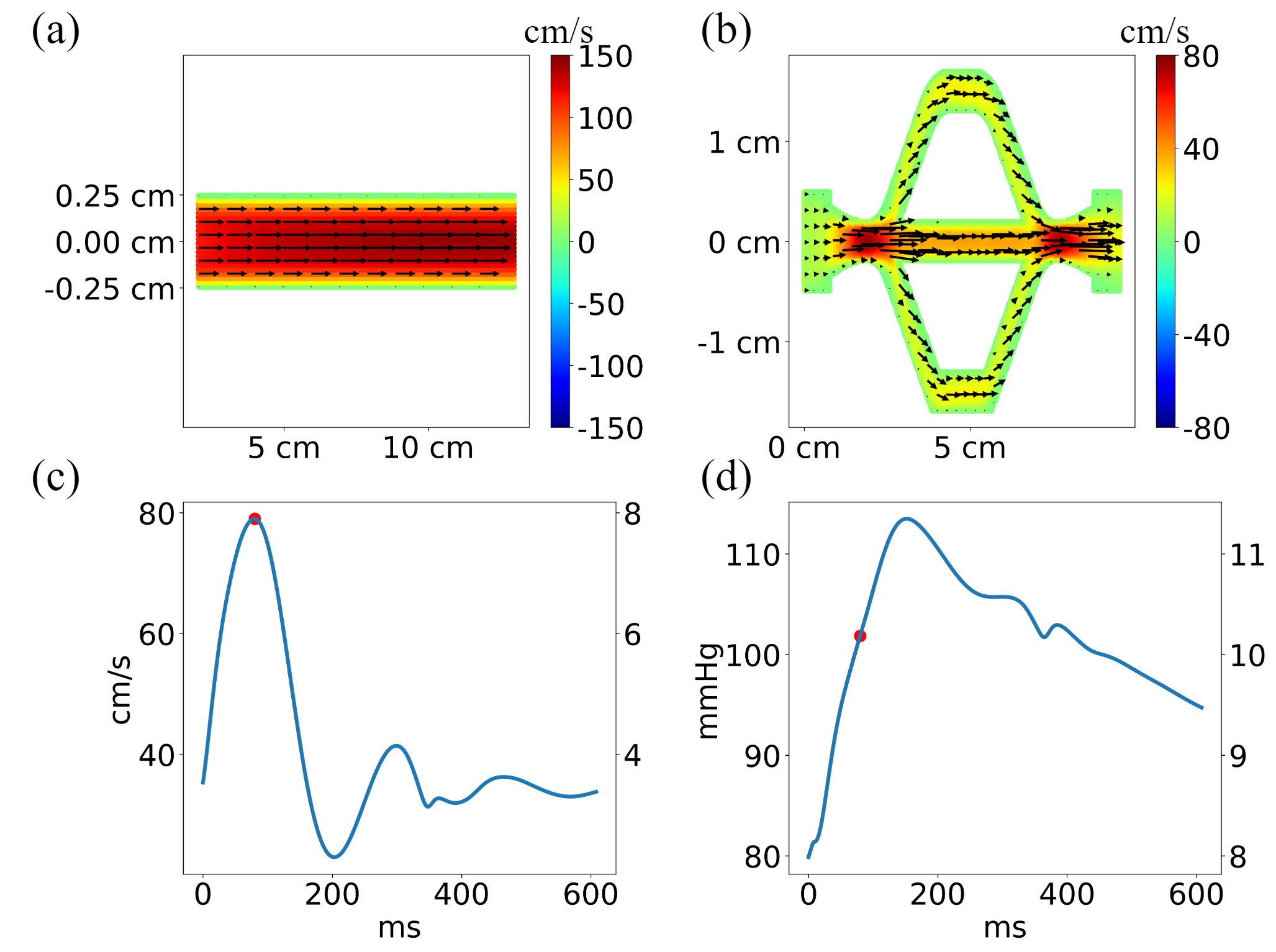}
\caption{Illustration of blood flow (only the lateral (i.e., horizontal) velocity component displayed here) in (a) a single-branch vessel and (b) a three-branch vessel with prescribed temporal profiles of the (c) flow velocity at the inlet and (d) blood pressure at the outlet. Black arrows in (a) and (b) represent blood flow velocity vectors.}
\label{fig:2}
\end{figure}
\subsection{Fluid-structure Interaction (FSI) Simulation}

We first evaluated our proposed methods on computer simulations of blood flow to demonstrate the feasibility of SeqPINN.  It is important to show that SeqPINN can solve Navier-Stokes equations under a pulsatile fluid flow as shown in Fig. \ref{fig:2}c given its relevance in biomedical applications. Fluid-structure interaction simulations, which better describe blood dynamics in a compliant blood vessel than CFD, were conducted \cite{dong2020walled} using Comsol 5.5 software (Comsol Inc. Burlington, MA, USA) and adopted as the ground truth. We designed a single-branch (Fig. \ref{fig:2}a) and a three-branch (Fig. \ref{fig:2}b) blood vessel to mimic the common segment of a major artery and branched structure, respectively. In building Comsol simulations, a surrounding medium as a deformable solid structure and its mechanical properties were predefined. The fluid properties were set close to real human blood, whose density $\rho$ was 1037 $kg/m^3$ and dynamic viscosity $\mu$ was 4.1 $mPa*s$. The Reynolds number of the blood flow in the single-branch blood vessel and three-branch blood vessel were $Re = \frac{\rho V_{avg}D}{\mu}=\frac{1037 kg/m^3*82.8cm/s*5mm}{4.1mPa*s}=1047.1$ and $Re = \frac{\rho V_{avg}D}{\mu}=\frac{1037 kg/m^3*18.7cm/s*10mm}{4.1mPa*s}=473$, respectively, where $V_{avg}$ is the average fluid velocity, and $D$ is the entrance diameter. The inlet flow velocity and outlet pressure used in the FSI simulation of the single-branch case are shown in Fig. \ref{fig:2}c and Fig. \ref{fig:2}d. In the single-branched vessel, 6,000 spatial collocation points were used. In the three-branch vessel, 25,000 spatial collocation points were used. 


 \begin{figure}[!t]
\centering
\includegraphics[width=0.9\linewidth]{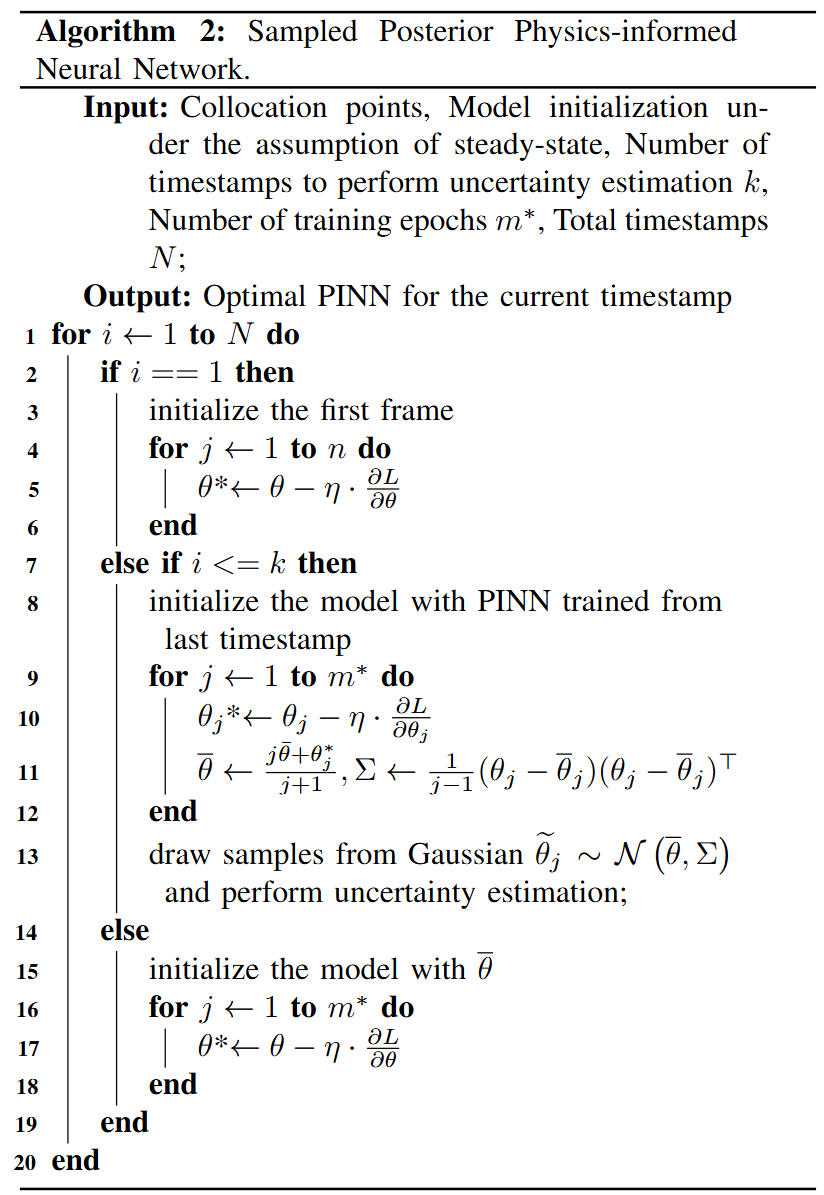}
\label{alg2}
\end{figure}

SeqPINN and SP-PINN were both trained on Nvidia GeForce RTX 3090. The training of SeqPINN and SP-PINN contained two stages: initialization and adaption. During initialization of SeqPINN, the model was trained for 3000 epochs using Adam optimizer. The learning rate was set to 1e-3 for 2000 epochs and 5e-4 for 1000 epochs. The batch size was 1024 in the single-branch vessel and 2048 in the three-branch vessel. The neural network consisted of 8 fully connected layers with 150 neurons in each layer. During adaption of SeqPINN, each timestamp was initialized with a pre-trained model from the previous timestamp and then trained for 30 epochs. The learning rate was set to 5e-4, following the end of training in the initialization stage. Batch size was kept at 1024 for the single-branch case and 2048 for the three-branch case. An identical initialization stage was performed for SP-PINN. Then, all timestamps were initialized with an averaged SGD solution during the adaption stage and thus trained in parallel. In our experiments, since the batch sizes (1024 and 2048) were small, full power of a single GPU could not be employed when a single model was trained. Thus, we utilized Multi-Process Service (MPS) provided by Nvidia to maximize the efficiency of GPU. To assess the accuracy and efficiency of our methods, we also trained Vanilla PINN and X-PINN\cite{CiCP-28-2002} with attention mechanism using the simulation data. 

\subsection{Vessel-mimicking phantom experiments}
We followed the protocol in our previous paper \cite{dong2020walled} to make a three-branch vessel mimicking phantom and validated the feasibility of SeqPINN and SP-PINN. A blood-mimicking fluid \cite{ramnarine1998validation} was made and circulated in the phantom at a constant flow rate of 2 ml/s controlled by a flow pump (BRAND). The training data was acquired using a Vantage 256 system (Verasonics Inc. Kirkland, WA) and an L11-4v probe (128 elements, Verasonics Inc. Kirkland, WA) operating at the center frequency of 6.25 MHz. Ultrafast image sequences were realized based on coherent plane wave compounding (CPWC) with three steered plane waves ($-10\degree,0\degree,10\degree$) at the compounded rate of 3000 frames per second (fps) \cite{tanter2014ultrafast}. 

\subsection{Evaluation metrics}
Model performance was evaluated in two aspects: accuracy and training efficiency. Root Mean Square Error (RMSE) across time and the relative error were both used to evaluate the model accuracy. RMSE is defined as $RMSE = \sqrt{\frac{1}{N}\sum_{i=1}^{N}(\hat{u_i}-u_i)^2}$ and relative error is defined as $\frac{1}{N}\sum_{i=1}^{N}|(\hat{u_i}-u_i)/max(u_i)|$, 
where $\hat{u_i}$ is the ground truth velocity, $u_i$ is the predicted velocity, $i$ represents index of collocation points, and $N$ is the total number of collocation points. 
In addition to training time, the ratio of training time to RMSE improvement is hereby proposed to evaluate training efficiency, which is central to real-time ultrafast ultrasound blood flow imaging. A large ratio indicates a huge sacrifice in training time for a small increase in RMSE, and this is not desired. 
\section{Results} \label{III}
\subsection{In Silico study}

\subsubsection{Single-branch Vessel}

\begin{figure*}[!t]
\centering
\captionsetup{justification=centering}
{\includegraphics[width=0.88\linewidth]{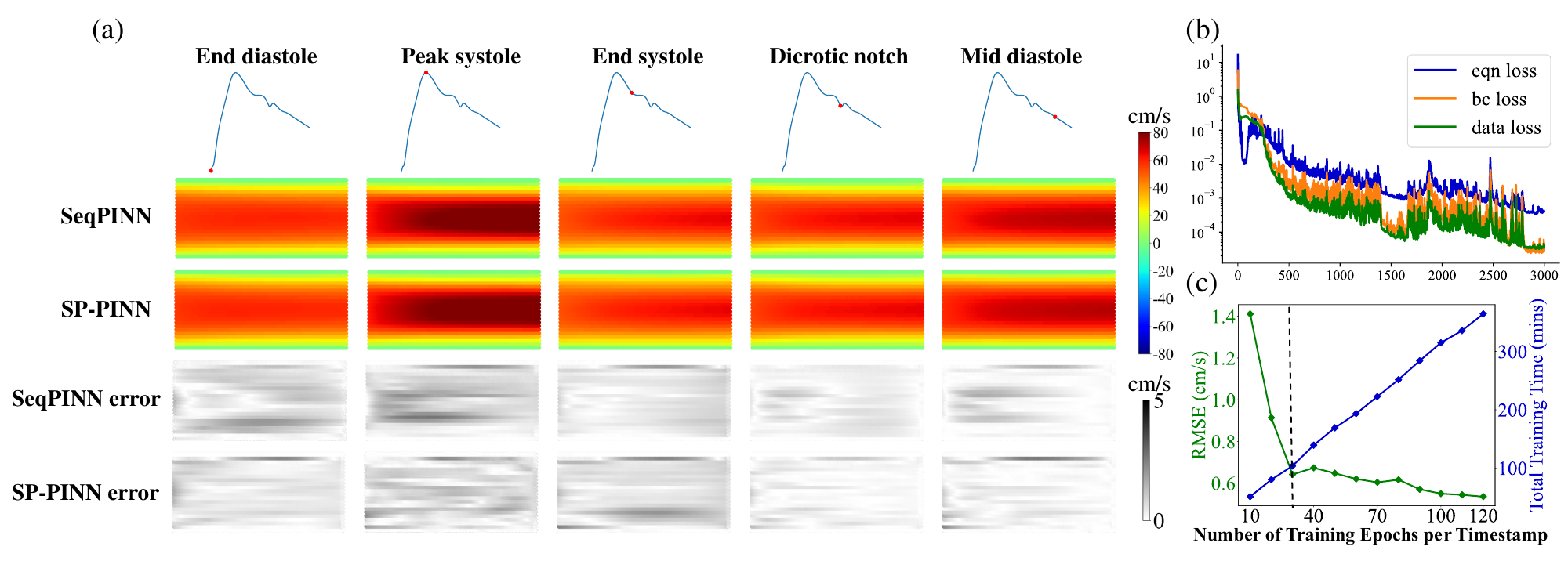}}
\caption{Learning the full fluid velocity field from sparse measurements on a single-branch vessel. (a) SeqPINN and SP-PINN predictions and error maps of the velocity fields at five representative phases in a cardiac cycle: \textit{End diastole, Peak systole, End systole, Dicrotic notch, Mid diastole}. Error maps show absolute errors. (b) Convergence of loss in the initialization of SeqPINN and SP-PINN. (c) Comparison of SeqPINN training time and accuracy as the number of training epochs per frame, $m$, increases in the simulated single-branch vessel case.}
\label{1boverview}
\end{figure*}

We first consider a single-branch vessel with 608 timestamps covering a cardiac cycle. Fig. \ref{1boverview}(a) shows the prediction and error maps from SeqPINN and SP-PINN for five important moments during a cardiac cycle: \textit{End diastole, Peak systole, End systole, Dicrotic notch}, and \textit{Mid diastole}. The ability to solve blood flow velocities during key phases in a cardiac cycle is of high clinical value. Fig. \ref{1boverview}(b) demonstrated the convergence of loss functions at the initialization stage for both SeqPINN and SP-PINN. The number of training epochs needed for subsequent timestamps after initialization, denoted as $m$, is an important parameter to be tuned in SeqPINN and SP-PINN to balance the trade-off between accuracy and efficiency. Fig. \ref{1boverview}(c) explained that an optimal $m*$ existed. In addition to model accuracy and efficiency, the convergence of SeqPINN and SP-PINN for timestamps after initialization was proven by the small standard deviations in RMSEs and relative errors of five independent runs as summarized in \cref{tbl:1}. In the single-branch vessel case \cref{tbl:1}, SeqPINN outperformed Vanilla PINN by 66\% in the RMSE value and its training time was 9.65 times shorter than Vanilla PINN; SP-PINN surpassed Vanilla PINN by 56\% in the RSME value and was 23.7 times faster than Vanilla PINN. Both SeqPINN and SP-PINN had a relative error of less than 1\%, thereby showcasing an excellent match between FSI simulation outputs and SeqpINN and SP-PINN predictions. 
  
\begin{figure*}[!t]
\centering
\captionsetup{justification=centering}
{\includegraphics[width=0.88\linewidth]{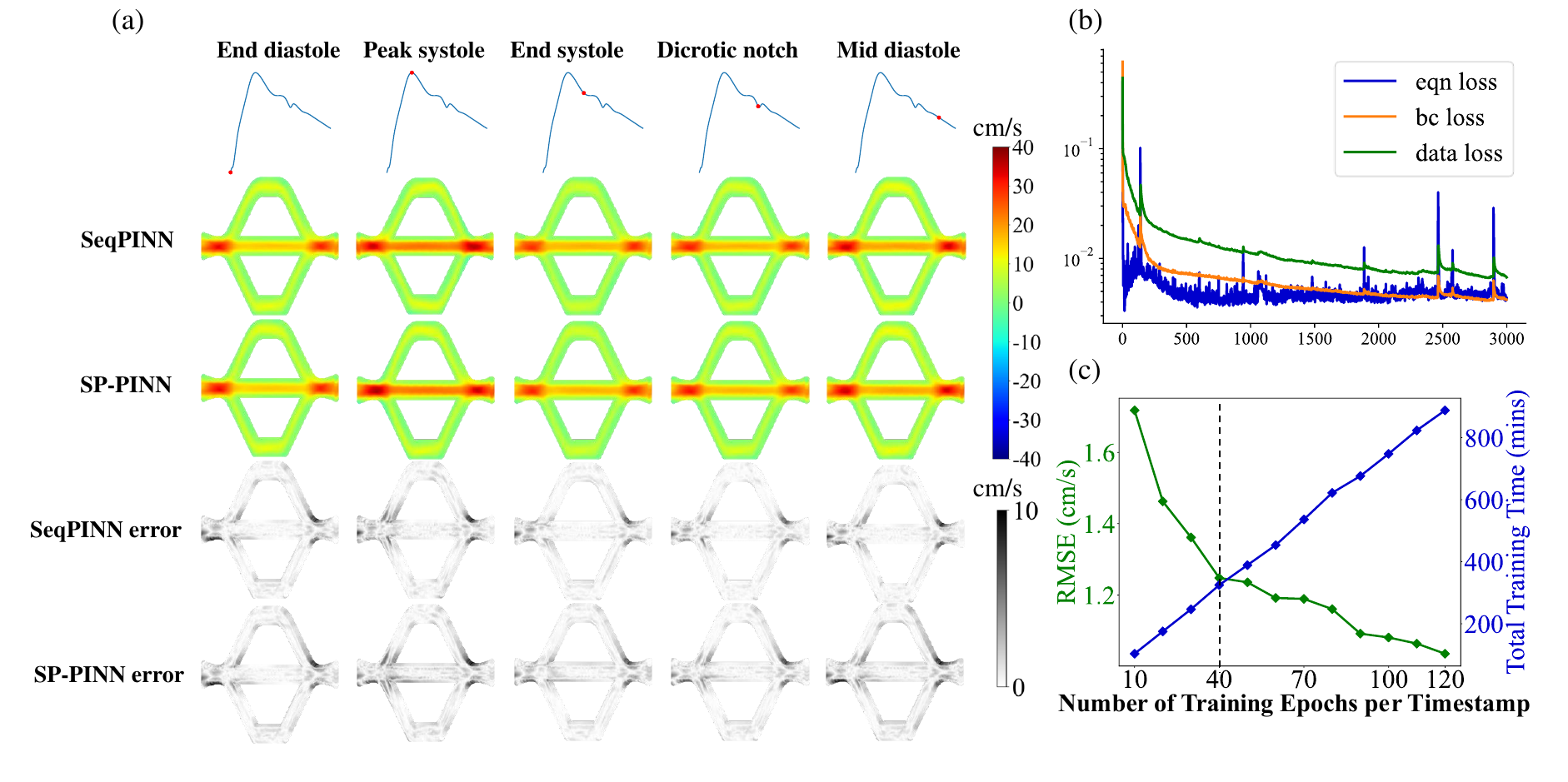}}
\caption{Learning the full velocity field from sparse measurements on a three-branch vessel. (a) SeqPINN and SP-PINN predictions and error maps of the velocity fields at five representative phases in a cardiac cycle: \textit{End diastole, Peak systole, End systole, Dicrotic notch, Mid diastole}. Error maps show the absolute errors. (b) Convergence of loss in the initialization of SeqPINN and SP-PINN. (c) Comparison of SeqPINN training time and accuracy as the number of training epochs per frame $m$ increases in the simulated three-branch vessel case.}
\label{3boverview}
\end{figure*}
\subsubsection{Three-branch Vessel}
Fig. \ref{3boverview}(a) shows the prediction and error maps of SeqPINN and SP-PINN on a trifurcated vessel, which had a more complex geometry than the single-branch one. Fig. \ref{3boverview}(b) depicted the convergence of loss functions at the initialization stage. It is worth noting that the optimal number of epochs $m*$ was 40 for the trifurcate vessel simulation case as shown in Fig. \ref{3boverview}(c). The RMSE of SeqPINN was 1.35 cm/s, which was 28\% lower than that of Vanilla PINN. The training speed of SeqPINN was 8.29 times faster than that of Vanilla PINN. The performance of SP-PINN was 13\% more accurate than Vanilla PINN, while the training time with SP-PINN was 29.6 times faster than Vanilla PINN.

\begin{figure}[!t]
\centering
\captionsetup{justification=centering}
{\includegraphics[width=0.9\linewidth]{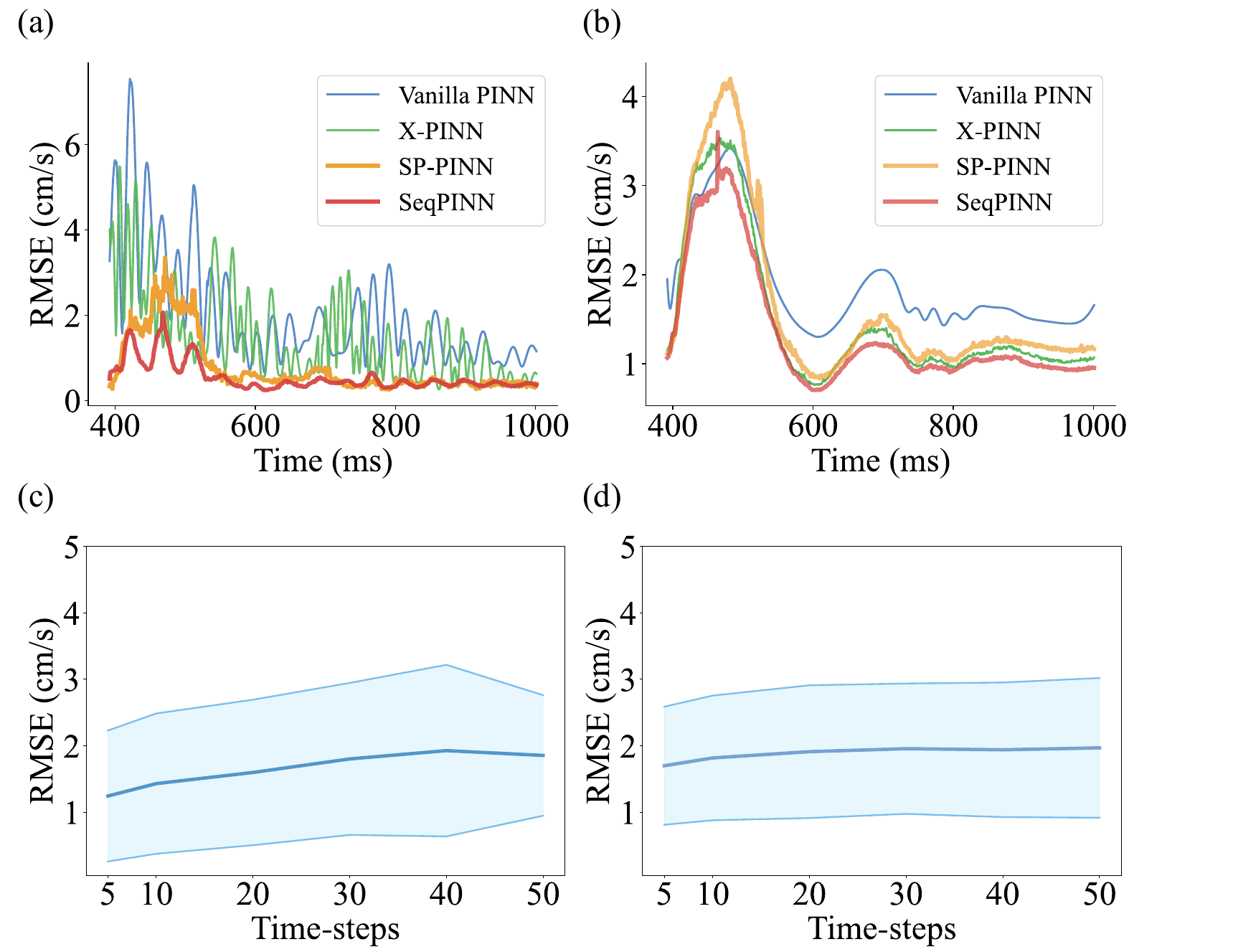}}
\caption{Comparison of RMSEs among Vanilla PINN, X-PINN, SeqPINN, and SP-PINN across 608 timestamps in the simulated (a) single-branch and (b) three-branch vessels. (c) and (d) show the comparison of accuracy and time-steps for SeqPINN on the simulated single-branch and three-branch vessels, respectively. Mean and standard deviation were calculated using RMSEs at each timestamp. Time-step sizes of 5, 10, 20, 30, 40, and 50 ms were tested.}
\label{results}
\end{figure}

Fig. \ref{results}(a) and Fig. \ref{results}(b) compared Vanilla PINN, X-PINN SeqPINN, and SP-PINN across all timestamps. Overall, the temporal profiles of the RMSE values followed the same pattern as the velocity profile. For the single-branch vessel (Fig. \ref{results}(a)), SeqPINN and SP-PINN performed similarly, while Vanilla PINN and X-PINN were more oscillating across time. X-PINN divided the computational domain into sub-domains and achieved slightly better performance than Vanilla PINN in accuracy and efficiency. SeqPINN was robust against a large time-step size (Fig. \ref{results}(c) and Fig. \ref{results}(d)). SeqPINN exhibited a strong transfer learning ability, with slightly degraded performance in accuracy as the step size increased. Note that the number of training epochs $m^{*}$ was kept at 30. The success of SeqPINN was based on a reasonable initialization for the current timestamp. Clearly, the initialization timestamp could be seen as not only a pre-trained model for the next timestamp but also a pre-trained model for all timestamps, hereby building a foundation for SP-PINN. 
\subsection{Phantom study}
\begin{figure}[!t]
\centering
\includegraphics[width=0.9\linewidth]{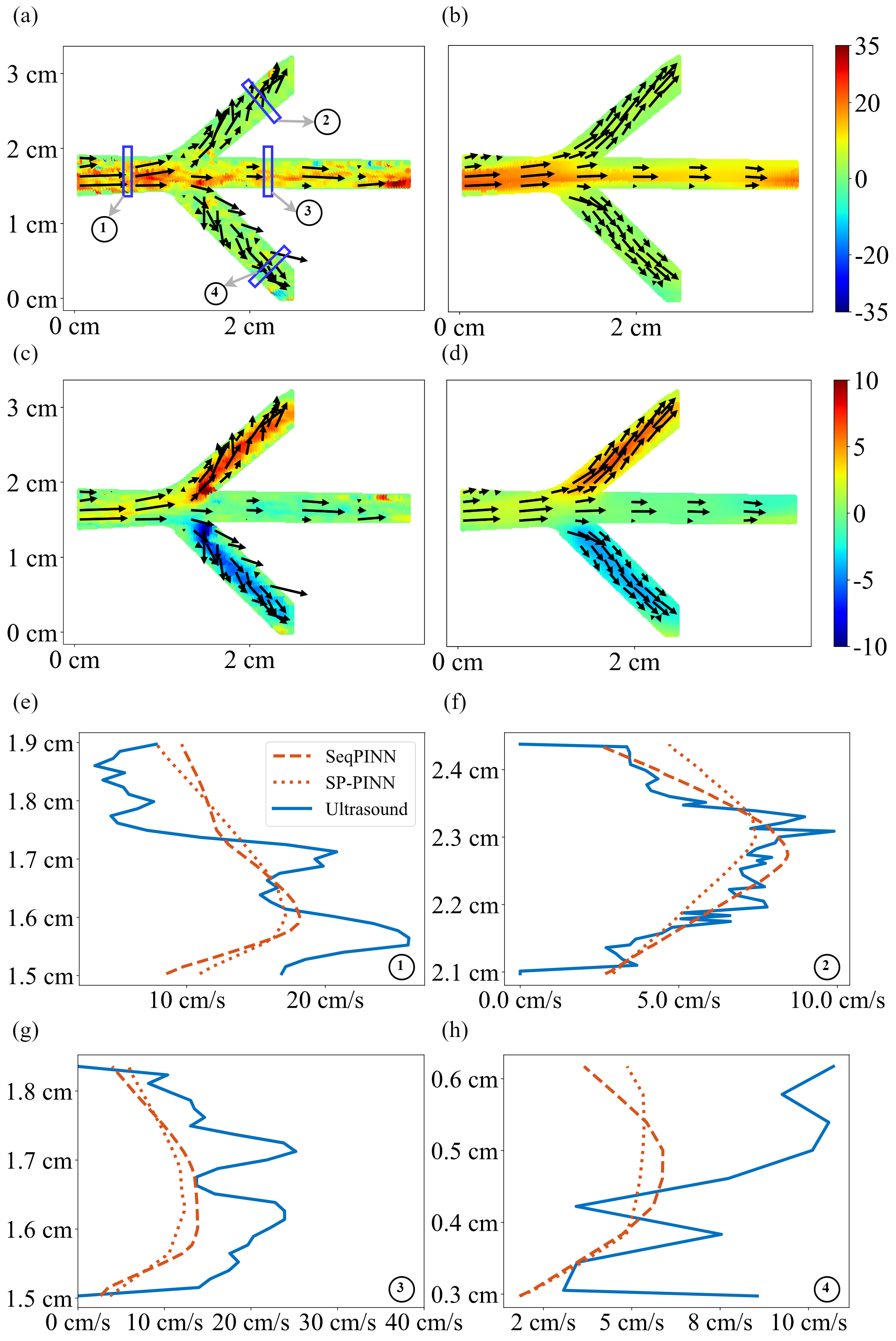}
\caption{Comparison of lateral (a) and axial (c) velocity maps obtained by a conventional ultrafast Doppler ultrasound method and their corresponding SeqPINN-regularized lateral (b) and axial (d) velocity maps. (e), (f), (g), and (h) are the velocity profiles over the cross-sections at locations 1, 2, 3, and 4 labeled in (a), respectively.}
\label{fig:4}
\end{figure}

The advantage of using PINN to regularize ultrafast ultrasound blood flow velocity estimates was demonstrated in vessel-mimicking phantom experiments. Fig. \ref{fig:4} shows estimated lateral and axial velocity maps together with flow vectors by a conventional ultrafast Doppler ultrasound technique and PINN-regularized results. PINN-regularized velocity maps were more spatially-continuous and exhibited parabolic velocity profiles across the lumen given a well-developed fluid flow scenario. Velocity profiles over corss-sections by SeqPINN and Sp-PINN agreed with each other.


\begin{table}
\caption{Comparison among FSI simulation, Vanilla PINN, X-PINN, SeqPINN, and SP-PINN on the single-branch vessel simulation across the entire computational domain. Mean and standard deviation (mean $\pm$ STD) are calculated based on five random runs.}
\centering
\begin{tabular}{@{} L L L L L @{} >{\kern\tabcolsep}l @{}} \toprule
 & \emph{RMSE (cm/s)} & \emph{Relative error \%}&\emph{Time (mins)}\\\midrule
 FSI   & -  & -  & 936.2     \\
 Vanilla PINN   & 1.85 $\pm$ 0.25 & 1.41 $\pm$ 0.12  & 965.23     \\ 
 X-PINN   & 1.59 $\pm$ 0.31 & 1.29 $\pm$ 0.15  & 748.52     \\ 
 SeqPINN  & \textbf{0.63 $\pm$ 0.09} & \textbf{0.38 $\pm$ 0.07}  & 100.05  \\ 
 SP-PINN  & 0.81 $\pm$ 0.12 & 0.43 $\pm$ 0.13 & \textbf{40.77}  \\\bottomrule
 \hline
\end{tabular} \label{tbl:1}
\end{table}

\begin{table}
\centering
\caption{Comparison among FSI simulation, Vanilla PINN, X-PINN, SeqPINN, and SP-PINN on the three-branch vessel simulation across the entire computational domain. Mean and standard deviation (mean $\pm$ STD) are calculated based on five random runs.}
\begin{tabular}{@{} L L L L @{} >{\kern\tabcolsep}l @{}}    \toprule
 & \emph{RMSE (cm/s)} & \emph{Relative Error \%} & \emph{ Time (mins)}   \\\midrule
  FSI   & -  & -  & 1448.43     \\
 Vanilla PINN   & 1.88 $\pm$ 0.18 & 3.52 $\pm$ 0.21  & 1788.25    \\ 
 X-PINN   & 1.77 $\pm$ 0.27 & 3.23 $\pm$ 0.17  & 1532.96     \\ 
 SeqPINN  & \textbf{1.35 $\pm$ 0.14} & \textbf{2.21 $\pm$ 0.16}  & 326.39  \\ 
 SP-PINN  & 1.63 $\pm$ 0.08 & 2.52 $\pm$ 0.17  &  \textbf{91.57} \\\bottomrule
 \hline
\end{tabular} \label{tbl:2}
\end{table}


\section{Discussion} \label{IV}
Developing a framework for fast training of PINN that incorporates a large number of timestamps in ultrafast ultrasound blood flow imaging is crucial for deciphering hemodynamics in arteries \textit{in vivo}. To tackle algorithmic challenges and physical limitations pertinent to Doppler-based and speckle-tracking-based velocimetry \cite{jensen2016ultrasound}, we proposed SeqPINN and SP-PINN under the assumption of the infinitesimal time-step to invoke physics into blood flow velocity estimation. Our study demonstrated that predictions of SeqPINN and SP-PINN agreed quantitatively with the ground truth in the FSI simulation study. We also tested the feasibility of our methods on phantom experiments. Our training framework significantly expedited the training process of PINN, exhibiting a potential for real-time training and clinical applications of PINN.

We demonstrated that SeqPINN was more computationally efficient by reducing the input dimension. Instead of training PINN to solve time-dependent Navier-Stokes equations, solving steady-state Navier-Stokes equations only requires (x,y) pairs as the input at a single timestamp. Assuming a case containing $n$ pairs of (x,y) and $N$ timestamps in total, the number of training points in an epoch for Vanilla PINN is $n\cdot N$, while that for SeqPINN is $n$. Thus, training one epoch of SeqPINN is always $N$ times more computationally efficient than PINN. Overall, training SeqPINN for 30 epochs over $N$ timestamps requires running $(30\cdot n)\cdot N$ data through the neural network, which is equivalent to training PINN, which requires $30\cdot (n\cdot N)$ points for 30 epochs. With the distinct advantages of pre-trained models and OTTA, SeqPINN could reach a solution in 30 epochs, while Vanilla PINN required 1200 epochs in our experiments. Similarly, with TTBA, SP-PINN reached a solution in 1 epoch after initialization. In terms of memory usage, Vanilla PINN requires $N$ times more copies of the collocation points into one optimization task, which may also cause GPU memory explosion when $N$ is large as in ultrafast ultrasound. In contrast, SeqPINN needs $N$ times more space for storage of PINN models. In other words, SeqPINN trades the number of models (storage space) for training speed. Fortunately, individual SeqPINN models in this study were very small in size (about 461 kB), and it was not necessary to store them on GPU. The trade-off between storage space and training speed is of high value in efficient training of PINN. Overall, we suggest SeqPINN in biomedical applications because accuracy outweighs the training speed. In other fields of research where the training speed is crucial, SP-PINN is a preferred option. Interested readers may refer to supplementary videos for fully-recovered blood flow velocity maps in the entire computational domain by SeqPINN and SP-PINN. For phantom experiments, blood flow velocity fields regularized by SeqPINN and SP-PINN as shown in Fig. \ref{fig:4} clearly exhibited a laminar pattern in all four locations, whereas original ultrasound estimates were noisy and less spatially continuous. The fact that the Doppler ultrasound technique adopted in this study can only estimate the axial velocity accurately provides a strategy for selecting sparse points as data loss in SeqPINN and SP-PINN training.

In SeqPINN, increasing $m$ by one for every timestamp resulted in $t$ more training epochs in total, with $t$ being the number of timestamps. Our results showed that the optimal number of training epochs $m^{*}$ existed to achieve satisfactory model accuracy. RMSE values initially decreased drastically and almost linearly as the number of training epochs increased for each timestamp until 30 epochs per timestamp, but then the decrease in RMSE values slowed down as each timestamp was trained for more epochs. Meanwhile, training time grew linearly as the number of training epochs for each timestamp increased. A diminishing return on the reduction in RMSE values was observed for every 10 more training epochs when $m$ was greater than 30. When training against a large time-step size as in Figs. \ref{results}(c)-(d), more training epochs are needed. 

\begin{figure}[!t]
\centering
\includegraphics[width=0.9\linewidth]{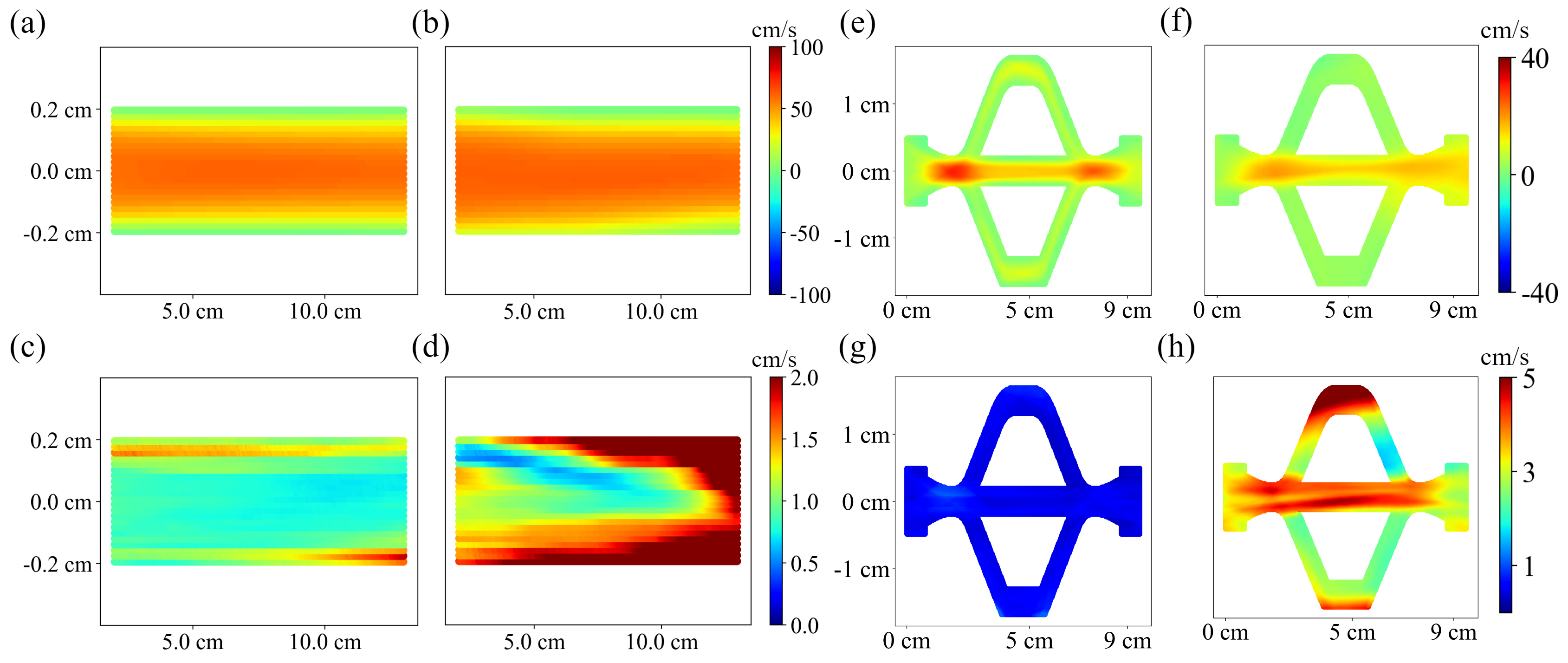}
\caption{Uncertainty estimation in the single-branch and three-branch vessels. (a) and (e) are samples from good estimation of the posterior distribution. (b) and (f) are samples from bad estimation of the posterior distribution. (c) and (g) are standard deviations (std) of 30 samples from the approximation of the posterior distribution corresponding to (a) and (e), respectively. Low std indicates that the initialization is stable. (d) and (h) are standard deviations (std) of 30 samples from the approximation of the posterior distribution corresponding to (b) and (f), respectively. High std indicates that the initialization is not stable.  }
\label{fig:8}
\end{figure}

SP-PINN was built upon the implementation of SeqPINN and an Bayesian approach. A well trained solution for the steady-state Navier-Stokes equations followed by a constant SGD training scheme produced a stable posterior Gaussian distribution of PINN, and the mean of the distribution shall have good generalizability across all timestamps. In SP-PINN, generalizability was assessed by an uncertainty map produced by sampling from the posterior distribution and performing Bayesian model averaging. Fig. \ref{fig:8} illustrates uncertainty estimation in single-branch and three-branch simulation cases. Note that in the single-branch vessel case, good initialization and bad initialization were similar visually. It was difficult to distinguish a bad initialization from a good initialization. Therefore, an uncertainty estimation was indispensable. In our experiments, the uncertainty index was monitored once every 500 training epochs. The size of the training dataset $k$ leveraged the training speed and training stability. When dealing with a pulsatile velocity field using a basic ultrafast Doppler ultrasound technique in this study, $k$ was much smaller than the total number of timestamps $N$. The lack of the global insight biased any efforts towards an initialization that generalized well to all timestamps. The Bayesian approach mitigated the bias with an informative prior belief and Bayesian marginalization. Practically, $k$ was fine-tuned in choices of {5, 10, 15, 20}. We found that using the average of 15 subsequent timestamps led to the best accuracy. The idea of averaging constant SGD solutions of consecutive timestamps was a cheap and well-justified way to approximate a Bayesian posterior distribution while ensuring accuracy for subsequent timestamps. Since we used the same initialization for all timestamps, SP-PINN was not affected by the step size in time. 
Few study limitations remain. First, that a high flow velocity caused difficulty in the optimization of PINN was observed in Vanilla PINN, X-PINN, SeqPINN, and SP-PINN. We speculated that high frequency functions due to the fast moving of blood flow was the key factor. Previous studies suggested that assigning different weights to terms in loss functions could alleviate the issue \cite{wang2022and}. Second, the branched structure posed a challenge for SeqPINN and SP-PINN in its ability to solve. Thus, the inlet velocity and outlet pressure were purposely preset to be 10 times smaller in the three-branch vessel than those used in the single-branch vessel. In our follow-up work, we will address the accuracy issue with branched structure. Third, we expedited the training of PINN by a large margin, but real-time training has not been achieved yet. The bottleneck in training of PINN is hard to break with current design of GPU. A special design of hardware-accelerated computation seems indispensable for real-time training of SeqPINN.   
\section{Conclusion}

In this paper, we developed SeqPINN with steady-state Navier-Stokes equations to accelerate the training of PINN, and SP-PINN to conduct parallel training of SeqPINN toward real-time generalization. The success of SeqPINN and SP-PINN attribute to the joint use of pre-trained model and TTA. Fast training of PINN is imperative in many areas, such as ultrafast ultrasound blood flow imaging, which aims to depict complex blood flow dynamics through thousands of frames per second. We view SeqPINN as the foundation towards real-time training of physics-informed learning for solving Navier-Stokes equations. The novel training framework of SeqPINN significantly reduces the training time, while achieving superior performance in both accuracy and efficiency compared with current implementation of PINN. SeqPINN is a generic algorithm that can incorporate various techniques mentioned in related work. For example, the design of weighted loss and calculation of derivatives can potentially be built on top of SeqPINN. SP-PINN with uncertainty estimation is a reliable way to initialize a generalizable model that can train SeqPINN in parallel. Meta-learning based approaches can also be adopted to search for a good initialization along the time dimension. The success of SeqPINN is built on the foundation of stead-state Navier-Stokes equations. This paper is envisioned to stimulate more future research on steady-state PDEs, particularly in biomedical applications.


\bibliographystyle{IEEEtran}
\bibliography{main.bib}

\end{document}